\documentclass[aps,prl,twocolumn,showpacs,amsmath,floatfix,citeautoscript]{revtex4}
\usepackage{graphicx}
\usepackage{dcolumn}

\begin{document}


\title{
Modeling functional piezoelectricity in perovskite superlattices with competing instabilities}

\author{Charles Swartz and Xifan Wu}
\email{xifanwu@temple.edu}
\affiliation{Department of Physics, Temple University, Philadelphia, PA 19122, USA}
\affiliation{Institute for Computational Molecular Science, Temple University, Philadelphia, PA 19122, USA}


\begin{abstract}
Based on the locality principle of insulating superlattices, we apply the 
method of Wu {\it et al} [Phys. Rev. Letter {\bf 101}
, 087610 (2008)] to the piezoelectric strains of individual layers under fixed
displacement field. For a superlattice of arbitrary stacking sequence an accurate 
model is acquired for predicting piezoelectricity. By applying the model in the superlattices where ferroelectric and 
antiferrodistortive modes are in competition, functional piezoelectricity can be achieved. A strong nonlinear effect
is observed and can be further engineered in the PbTiO$_3$/SrTiO$_3$ superlattice and an interface enhancement of piezoelectricity is found in the BaTiO$_3$/CaTiO$_3$ superlattice.
\end{abstract}

\pacs{77.22.-d, 77.22.Ej, 77.80.-e, 77.84.Lf}

\maketitle




\def\xwm#1{\marginpar{\small XW: #1}}
\def\scr{\scriptsize}

Multicomponent ABO$_3$ perovskite superlattices(SLs) provide a very promising way
to design novel materials with multifunctional properties for device application~\cite{Lee_Nature, Ghosez_Nature}.
Bridged by the interfaces, distinct instabilities belonging to individual bulk constituents
are in strong competition with these artificial materials. Functional properties 
such as polarization~\cite{Lee_Nature, Ghosez_Nature, Bousquet_PRB_2010, Wu_PRB_2011}, 
piezoelectric~\cite{Evans_PRL_2010, Evans_PRL_2008}, multiferroic~\cite{Fennie_PRL_2006, Rabe_PRL_2010, Scholom_Nature_2010}, 
and dielectric responses are 
found to be highly sensitive to the interactive instabilities, sometimes, resulting 
in unexpected enhanced functionalities.

Paraelectric(PE)/ferroelectric(FE) SLs with both antiferrodistortive(AFD)
and FE instabilities have attracted intense attention recently~\cite{Ghosez_Nature,Wu_PRB_2011, Rondinelli_AM_2011}. 
The zone-boundary nonpolar AFD mode
associated with the oxygen octahedral rotation and the zone-center polar FE mode are usually found
to be exclusive to each other in bulk perovskite. However in SLs of 
PbTiO$_3$(PT)/SrTiO$_3$(ST) and BaTiO$_3$(PT)/CaTiO$_3$(ST), first-principles 
calculations revealed that the AFD and FE can coexist with an interface 
reconstruction~\cite{Ghosez_Nature,Wu_PRB_2011}. In both cases, the competing AFD and FE modes at interfaces are
predicted to enhance the polarization, which is consistent with the experimental 
observation~\cite{Ghosez_Nature, Lee_APL_2009}.

Clearly, for these insulating PE/FE SLs, both ionic and electronic interface
effects should be well localized. The local electrostatic property will only be different 
from the bulk material within a few layers away from the interface under the fixed displacement
field. Thus the interface dipoles can be accurately described by a 
truncated cluster expansion(TCE) model developed by Wu {\it et al.}~\cite{Wu_PRL_2008}; 
in which the electronic states
of interfaces are represented by the maximally localized Wannier functions through a unitary transformation
from Bloch-like orbitals~\cite{MLWF, Giustino03, Wu_PRL_2006}.
Based on the above, SL design can be performed in the AFD/FE competing system, where the interface
is expected to increase the overall FE. For quite some time, the focus was to explore the 
functional polarization~\cite{Ghosez_Nature, Bousquet_PRB_2010, Wu_PRB_2011, Junquera_axiv, Lee_AM, Dawber_AM}. 
Piezoelectricity~\cite{Spaldin_PRL_2005,Wu_PRB_2005, Fu_Nature, Rabe_PRB_2009}, another important functionality, 
describes the coupling between polarization and strain. Its interface mechanism and SL
design rule have not yet been addressed in these intriguing systems with competing instabilities.

In this letter, we show that functional piezoelectricity can be designed in the 
SLs through the AFD and FE competition and its electric field dependence.
We further apply our TCE model to the piezoelectric strain of  
the SLs in a fixed displacement field. In combination with a similar treatment of 
Wannier based layer polarization, we arrive at an accurate modeling for predicting the piezoelectric 
tensor for an arbitrary sequence of SL. In bulk ST, we discover a strong nonlinear
piezoelectric response originating from the completely suppressed AFD  
in a large applied electric field. We use the model to demonstrate that the electric field of this 
nonlinear piezoelectricity can be reduced to a much smaller magnitude with an largely increased field tunability 
by changing the PT fraction in PT/ST SLs. With the model, we are able to systematically study the interface 
effect on the piezoelectric response in both BT/CT and PT/ST SLs. In BT/CT SLs, we 
find a novel interface enhancement of piezoelectricity.

The first-principles calculations are carried out in two series of SLs $n$PT/$m$ST and 
$n$BT/$m$CT both with AFD and FE instabilities. These include the bulk PT, ST, BT, CT and 
all period-four SLs($n+m=4$) stacked in the [001] direction. The fixed in-plane lattice constant 
is chosen as $a_0$=$7.275$\,bohr, the computed equilibrium lattice constant for bulk cubic SrTiO$_3$.
We assume the epitaxial growth of the SLs on SrTiO$_3$ which is consistent with the recent 
experiments~\cite{Ghosez_Nature, Evans_PRL_2010}. 
Both the FE mode along [001] and the AFD mode associated with oxygen octahedral TiO$_6$ rotation
around [001] axis have been explicitly included in the sub-lattice structure of $P4bm$ space-group symmetry.
To this end, a 40-atom tetragonal supercell is adopted with lattice vectors of length $\sqrt{2}a_0$ in
the [110] and [1$\bar 1$0] directions and [1$\bar 1$0] directions and $c \approx 4a_0$ along 
the [001] direction.

We use density-functional theory implemented in the LAUTREC code package \cite{Max_all} 
to perform structural relaxations and electron minimizations at fixed electric displacement fields.
Local-density approximation \cite{LDA_PW92} is adopted and
plane-wave calculations are implemented in the projector augmented-wave
framework \cite{PAW}. We used a plane-wave cutoff energy of 80\,Ry and a 4$\times$4$\times$1
Monkhorst-Pack $k$ mesh.

Our model for piezoelectricity starts from the decomposition of the SLs into
AO and BO$_2$ layers along [001] direction. For each fixed-D field, both Wannier-based layer
polarization $p_i(D)$~\cite{Wu_PRL_2006} and the layer height $h_i(D)$~\cite{Wu_PRB_2011}
(which is directly related to piezoelectric strain $\eta_i(D)$) of each individual layer are computed 
from the relaxed electronic and ionic structure respectively. Working with the constrained-$D$ field 
framework~\cite{Max_all}, we employ longitudinal boundary condition which limits the force constant matrix to short
range interaction of a few neighbors. 

\begin{figure}
\centering
\includegraphics[width=2.7in]{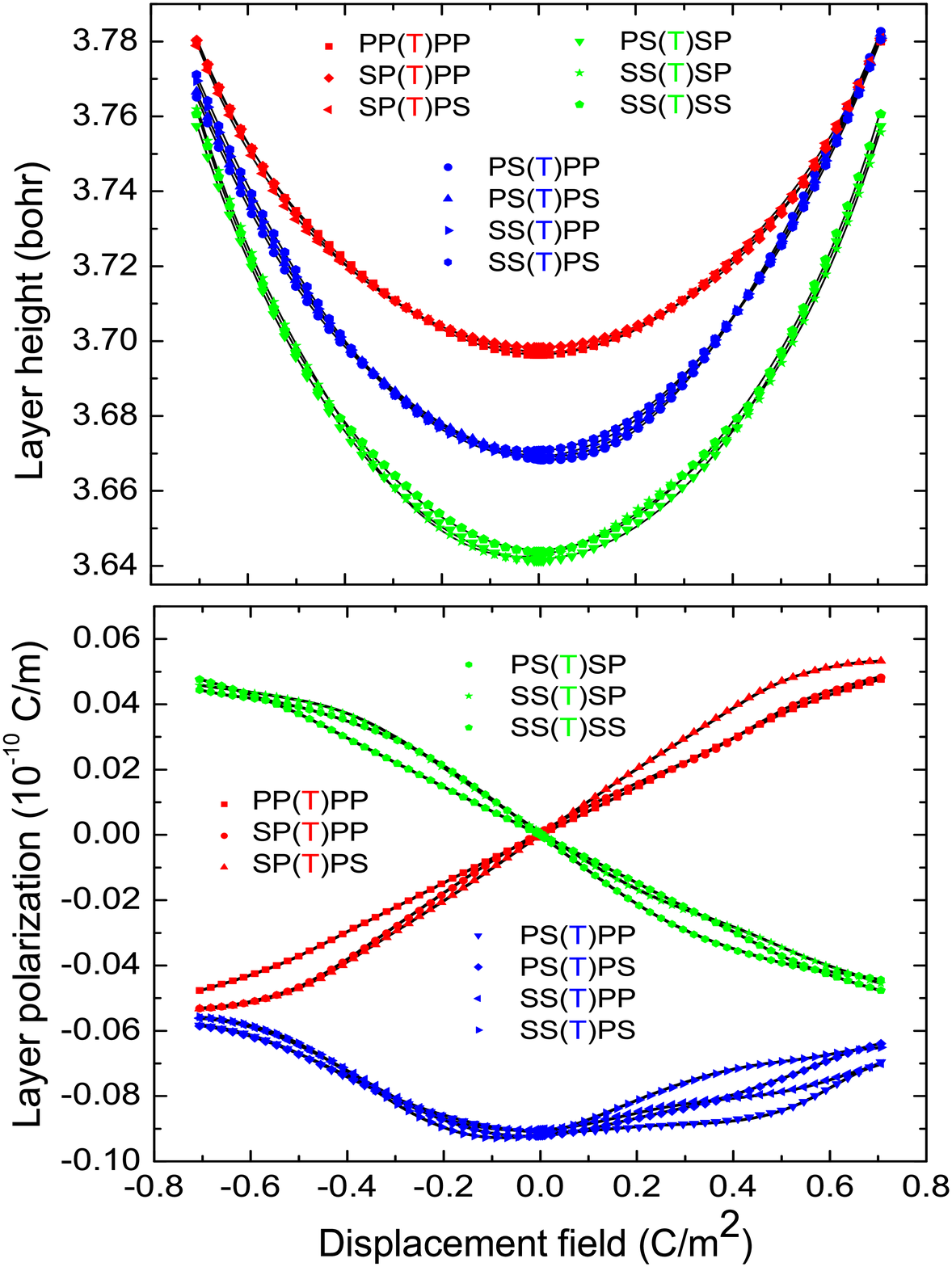}
\caption {(color online) Dependence of TiO$_2$(T) (a) layer height 
(b) layer polarization(relative to the average of the TiO$_2$ planes of bulk PbTiO$_3$ and SrTiO$_3$)
on chemical environment in PT/ST SLs.}
\label{fig1}
\end{figure}

In Fig.~1, we present the representative $h_j(D)$ and $p_j(D)$ curves in PT/ST SLs.
As expected, the locality principle is satisfied on not only the layer 
polarization~\cite{Wu_PRL_2008} but layer height.
As a result both $h_j(D)$ and $p_j(D)$ curves separate into clusters depending on the nearest neighbor 
chemical environment. In can be seen that $p_j(D)$ is mostly determined by the identity of nearest neighbors and
has a much weaker dependence on the second neighbors. Compared with $p_j(D)$, $h_j(D)$ shows a
similar behavior with an even stronger localization. A clear local inversion symmetry breaking is also 
present in $h_j(D)$ as well as $p_j(D)$.  
As an example, the inversion symmetry breaking introduced by the first neighbors has a large asymmetric behavior for the 
$h_j(D)$ of TiO$_2$ in the middle of the S(T)P sequence. This asymmetric behavior becomes much 
weaker when the local inversion symmetry breaking occurs only on the second neighboring layers in SS(T)SP.
In previous work~\cite{Wu_PRL_2008}, it has been shown that a TCE model accurately captures 
the dependence of $p_i(D)$ on its local compositional environments.
The similar locality of $h_i(D)$ indicates that $h_j(D)$(as well as  $p_j(D)$) can be accurately described by
the TSE as
\begin{eqnarray}
h_l (\{s\}) & = & J_0 
   + \sum_{i}\left(  J_{l,i} s_i + J_{l,i}^{\prime}s_i^2 \right) \cr
  & + & \sum_{ij} \left( J_{l,ij} s_is_j
   +  J_{l,ij}^{\prime}s_is_j^2
   +  J_{l,ij}^{\prime\prime}s_i^2s_j^2 \right) \cr
  & + & \sum_{ijk}J_{l,ijk} s_is_js_k + ... .
\label{eq:CE_general}
\end{eqnarray}
Our TCE model will include the cluster interaction of one particular layer with the neighboring 
layers up-to second nearest neighbors and only include up to two-body interaction terms.
As a result, all the truncated cluster terms are effectively included in the compositional environment 
of the period-four SLs. All the computed $p_j(D)$ and $h_j(D)$ will serve as a first-principles
database. The $J$ terms are D-dependent effective cluster interaction coefficients and are computed from the database.
The ``pseudo-spin'' variable is defined as $s_i = 1,0$ and is used to identity the $i$th AO layer as either PbO(BaO) or SrO(CaO) respectively.
The expressions for $h_{\rm AO}$, $h_{\rm TiO_2}$, $p_{\rm AO}$ and $p_{\rm TiO_2}$ with all the fitted 
ECIs will be given in the on-line supporting material~\cite{Support}. The total supercell lattice constant is obtained from
the sum of all the individual layer heights as $h(D)=\sum_ih_i(D)$ and the piezoelectric strain 
of the SL can be calculated by $\eta(D) = (h(D)-h(D=0))/h(D=0)$. 
The electric equation of state of $\eta$ as a function of electric field ${\cal E(D)}$ will be obtained by numerical inversion
The electric field is computed by ${\cal E}(P)=D(P)-4\pi P$, where the total polarization of the SL
is given by $P(D)=h(D)^{-1}\sum_{j}p_j(D)$. It is then straightforward to compute the piezoelectric coefficient
$d_{33}=\partial \eta_{33} / \partial {\cal E} $.

Here we want to stress that the modeling of the individual layer hight $h_j(D)$, 
instead of the total height $h(D)$ of the SL, is important for a more accurate piezoelectric model. 
This is because the piezoelectricity describes the derivative of the strain with respect to 
the electric field, which is sensitive to the accuracy of the strain model. Due to the short-range 
nature, the models of $p_j(D)$ and $h_j(D)$ can be used to compute $\eta(D)$ and ${\cal E(D)}$ 
for an arbitrary sequence of SL. This enables us to study the piezoelectricity resulting from the 
AFD/FE competition under an applied electric field or at the interfaces systematically.

\begin{figure}
\centering
\includegraphics[width=2.7in]{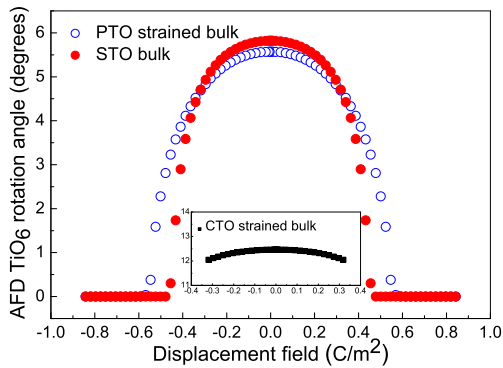}
\caption{\label{fig2} (Color online) TiO$_6$ rotation as a function of D field for TiO$_6$ octahedral 
in bulk ST, strained bulk PT and strained bulk CT(shown in insert).}
\end{figure}

It is well established that AFD and FE will compete with each other in bulk 
ABO$_3$~\cite{Rondinelli_AM_2011, Zhong_PRL_1994,Ghosez_PRL_2009}. Under an applied 
electric field, the FE will be strengthened by the coupling between the electric field and polar mode and
the SL will be further polarized along the field direction. It is thus expected that AFD will be 
suppressed with increased field. Indeed in Fig.~2, we can see that the octahedral rotation is largely 
suppressed with increased D magnitude in both bulk ST and strained bulk PT. Surprisingly at 
$D \sim 0.4~C/m^2$ and $D \sim 0.6 ~C/m^2$, the octahedral rotation will 
be completely suppressed resulting in a FE only phase. 
It is also important to notice that the TiO$_6$ rotation in bulk PT is {\it hidden} and can not be observed 
experimentally. This is because the AFD instability is completely suppressed in the metastable region
before the spontaneous polarization is reached ($D \simeq P =  0.8~C/m^2$). In the PT/ST SLs,
the spontaneous polarization will be reduced and the AFD rotation will be recovered in the PT fraction.
In contrast, octahedral rotations in strained bulk CT have a much weaker field dependence.
The TiO$_6$ rotation in CT is only slightly decreased from 13$^\circ$ 
at $D=0$ as shown in the inset of Fig.~2~\cite{Wu_PRB_2011}. 
This is consistent with the much stronger AFD than FE instability which results in the 
PE ground state for bulk CT. On the other hand BT shows no AFD rotation at all for the whole field 
range. This is expected from the fact that bulk BT 
highly resists AFD rotation with a robust FE ground state.

Since ST is PE at ground state, the disappearance of the AFD instability indicates a physical
phase transition driven by an applied electric field. During the phase transition, SL structure 
changes from AFD/FE into a FE only phase accompanied by a structure softening,
which is signaled by a strong nonlinear piezoelectric at finite electric field.
One can clearly see an additional peak of $d_{33}({\cal E})$ in bulk ST in Fig.~3. However, 
it occurs at a very high electric field(centered around 600 Mv/m). This is probably the reason 
why this phase transition has not been addressed yet in the literature.

\begin{figure}
\centering
\includegraphics[width=2.5in]{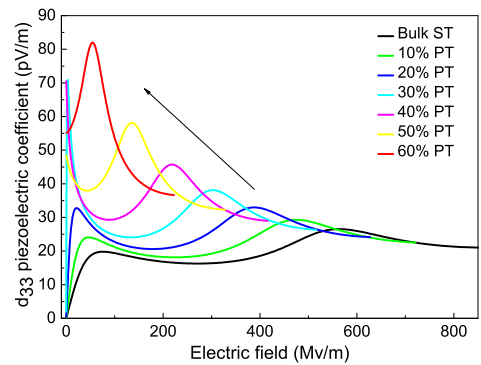}
\caption{\label{fig3} (Color online) Dependence of d$_{33}(\cal E)$ piezoelectric coefficient on 
PTO fraction in the $n$PT/$m$ST SLs.}
\end{figure}

Under the fixed-D field, both the FE and AFD ordering is short-ranged. So it should be understood that 
the D field is the fundamental variable that drives the phase transition at which the threshold D field 
is reached. Keeping the above concept in mind, we propose that D-field can be increased by a highly polarizable
PT component in the PT/ST SLs. As a result of the modified energetics of AFD and FE, the above 
phase transition, as well as the piezoelectric anomaly, can be observed at a much lower electric field.
We then use our developed model to test the above idea. We set up a $n$PT/$m$ST SL 
where $n+m=200$ and gradually increase the PT fraction from ST bulk $n=0$. 
The model prediction of the piezoelectric coefficient is presented in Fig.~3.
As expected, we can see that the SLs become more polarized as the PT fraction is increased.
The SL starts from a PE state and becomes FE after the PT fraction in the SL is larger than $\sim$~30\%.
As a result of increased FE ordering, the piezoelectric coefficient is generally increased with the anomalous increase 
corresponding to the phase transition in the ST component. Furthermore, the center of this anomaly is gradually shifted to lower electric field (centered at 55 mV/m for 60\% PT fraction). In having increased polarizability from a higher PT fraction in the SLs, 
the width of the piezoelectric anomaly also becomes smaller indicating the greatly increased tunability of applied electric
field.

AFD and FE competition at interfaces is also found to be of critical importance for its 
functional properties~\cite{Ghosez_Nature, Wu_PRB_2011}.
Here, focusing on the piezoelectricity, we can use our model to study the interface effect in both BT/CT and 
PT/ST systems. To the above end, we set up the SL model of $n$PT/$n$ST and $n$BT/$n$CT and 
gradually increase $n$ from dense interface limit ($n=1$) to the interface free limit ($n=\infty$).
The model prediction of the $d_{33}(\cal E)$ is presented in Fig.~4(the predicted spontaneous 
polarizations $P_s$ as a function of $n$ are also plotted in the inserts of Fig.~4(a) and (b)). 
The $d_{33}(\cal E)$ from directed first-principles are also given for comparison. 
One can see that our model is very accurate in reproducing the
first-principles results. In both systems the SL will have the largest $P_s$ at the dense interface limit
($P_s$= 0.187 C/m$^2$ for 1BT1CT and 0.327 C/m$^2$ for 1PT1ST). The $P_s$ will monotonically decrease until saturate 
at 0.21 C/m$^2$ and 0.14 C/m$^2$ for $n$PT/$n$ST and $n$BT/$n$CT in the interface-free limit.
This is consistent with the interface enhancement of polarization found in these materials.
Surprisingly, we see also an interface enhanced piezoelectricity in $n$BT/$n$CT where $d_{33}(\cal E)$ gains its 
maximum magnitude at $n=1$. In contrast the $n$PT/$n$ST SLs show the opposite effect, where the nonlinear effect 
corresponding to the ST phase transition starts to be observed when $n=3$.

\begin{figure}
\centering
\includegraphics[width=2.8in]{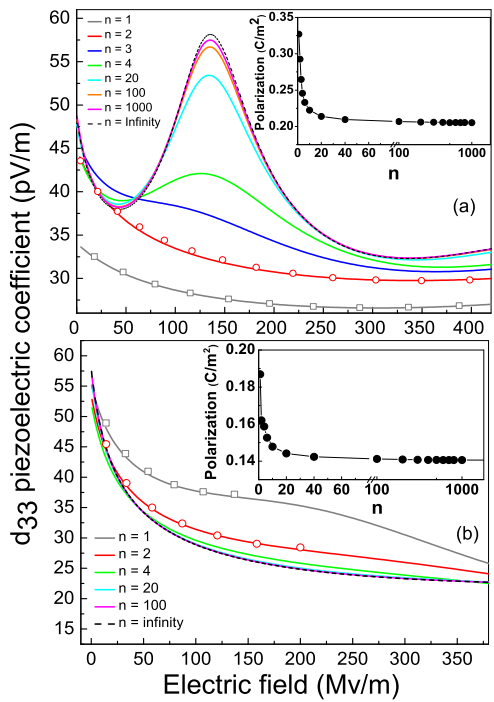}
\caption{\label{fig4} (Color online) Dependence of d$_{33}(\cal E)$ piezoelectric coefficient on the density 
of interfaces in (a) $n$PT/$n$ST SLs (b) $n$BT/$n$CT SLs.}
\end{figure}

The increased $d_{33}(\cal E)$ indicates the structural softening introduced by the interfaces.
In FE materials, the structural softening can be found when the system is approaching the FE/PE phase boundary 
e.g. with changing in-plane epitaxial strain~\cite{Fennie_PRL_2006}. As a result, the dielectric and piezoelectric responses diverge
in the vicinity of phase boundary. Indeed, we do see a similar interface dependence of the static dielectric response
(including the piezoelectric mediated component)~\cite{Support} confirming the 
additional softening of the polar mode due to the interfaces.
Strikingly, in BT/CT SLs the polarization and piezoelectric response can be both increased by the interface which is 
crucial for SLs design for multifunctional properties.

In summary, we have developed an accurate model that can the predict of piezoelectric coefficient for an
arbitrary sequence of SL using the first-principles results of short-period SLs only.
The predictive power of the model has been demonstrated in the PT/ST and BT/CT systems in 
which AFD and FE are in strong competition.
Functional piezoelectricity can be designed in the PT/ST and BT/CT SLs
under an applied electric field or a function of interface density.

\acknowledgments
XW thanks K. M. Rabe, A. M. Rappe, D. Vanderbilt and M. Stengel for useful
discussions. XW acknowledges the computational support by the National Science Foundation through 
TeraGrid resources provided by NICS under grant number [TG-DMR100121].

\end{document}